\documentclass[fleqn,usenatbib]{mnras}

\usepackage{newtxtext,newtxmath}

\usepackage[T1]{fontenc}
\usepackage{enumitem}
\usepackage{array}

\DeclareRobustCommand{\VAN}[3]{#2}
\let\VANthebibliography\thebibliography
\def\thebibliography{\DeclareRobustCommand{\VAN}[3]{##3}\VANthebibliography}

\usepackage{newtxtext, newtxmath}
\usepackage{graphicx}	
\usepackage{amsmath}	
\usepackage{orcidlink}

\usepackage{leftindex}	
\usepackage{amsmath}	
\usepackage{subcaption}

\usepackage{physics}

\usepackage[T1]{fontenc}
\usepackage{ae,aecompl}

\usepackage{amsmath,amsfonts,bm}

\def\eqref#1{equation~\ref{#1}}

\def\1{\bm{1}}

\def\vb{{\bm{b}}}

\DeclareMathAlphabet{\mathsfit}{\encodingdefault}{\sfdefault}{m}{sl}
\SetMathAlphabet{\mathsfit}{bold}{\encodingdefault}{\sfdefault}{bx}{n}

\usepackage{graphicx}	
\usepackage{amsmath}	
\usepackage{bm}

\newcommand{\bx}{\mathbf{x}}

\newcommand*\diff{\mathop{}\!\mathrm{d}}
\newcommand*\expm{\mathop{}\!\mathrm{exp}}

\usepackage{physics}

\def\btheta{\boldsymbol{\theta}}

\title[Geometric Deep Learning and Galaxies ]{Geometric deep learning  for galaxy-halo connection: a case study for galaxy intrinsic alignments}

\author[Jagvaral et al.]{
Yesukhei Jagvaral$^{1,2,4}$\thanks{E-mail: yjagvara@andrew.cmu.edu}\orcidlink{https://orcid.org/0000-0001-7068-7037},
Fran\c{c}ois Lanusse$^{3,4}$\orcidlink{0000-0001-7956-0542},
Rachel Mandelbaum$^{1,2}$\orcidlink{0000-0003-2271-1527}
\\
$^{1}$McWilliams Center for Cosmology and Astrophysics, Department of Physics, Carnegie Mellon University, Pittsburgh, PA 15213, USA\\
$^{2}$NSF AI Planning Institute for Data-Driven Discovery in Physics, Carnegie Mellon University, Pittsburgh, PA 15213, USA\\
$^{3}$AIM, CEA, CNRS, Universit\'e Paris-Saclay, Universit\'e Paris Diderot, Sorbonne Paris Cit\'e, F-91191 Gif-sur-Yvette, France\\
 $^{4}$ Simons Foundation, Flatiron Institute,  New York, NY 10010, USA \\
}

\date{Accepted XXX. Received YYY; in original form ZZZ}

\pubyear{2015}

\begin{document}
\label{firstpage}
\pagerange{\pageref{firstpage}--\pageref{lastpage}}
\maketitle

\begin{abstract}
Forthcoming cosmological imaging surveys, such as the Rubin Observatory LSST, require large-scale simulations encompassing realistic galaxy populations for a variety of scientific applications. Of particular concern is the phenomenon of intrinsic alignments (IA), whereby galaxies  orient themselves towards overdensities, potentially introducing significant systematic biases in weak gravitational lensing analyses if they are not properly modeled. Due to computational constraints, simulating the intricate details of galaxy formation and evolution relevant to IA across vast volumes is impractical. As an alternative, we propose a Deep Generative Model trained on the IllustrisTNG-100 simulation to sample 3D galaxy shapes and orientations to accurately reproduce intrinsic alignments along with correlated scalar features. We model the cosmic web as a set of graphs, each graph representing a halo with nodes representing the subhalos/galaxies. The architecture consists of a SO(3) $\times$ $\mathbb{R}^n$  diffusion generative model, for galaxy orientations and $n$ scalars, implemented with E(3) equivariant Graph Neural Networks that explicitly respect the Euclidean symmetries of our Universe.   The model is able to learn and predict  features such as galaxy orientations that are statistically consistent with the reference simulation. Notably, our model demonstrates the ability to jointly model Euclidean-valued  scalars (galaxy sizes, shapes, and colors) along with non-Euclidean valued SO(3) quantities (galaxy orientations) that are governed by highly complex galactic physics at non-linear scales. 
\end{abstract}

\begin{keywords}
methods: numerical – cosmology: theory – galaxies: statistics – galaxies: structure – gravitational lensing: weak
\end{keywords}







\section{Introduction}\label{intro}

Ongoing and future cosmological weak lensing surveys,   including the European Space Agency's Euclid Mission\footnote{ \url{https://www.euclid-ec.org/} }, the Vera C.\ Rubin Observatory Legacy
Survey of Space and Time\footnote{ \url{https://www.lsst.org/} }, and  the Nancy Grace Roman Space Telescope\footnote{ \url{https://roman.gsfc.nasa.gov/} } High Latitude Imaging Survey, 
  promise a wealth of data for precise cosmological parameter measurements \citep{Kilbinger}. 
  However, extracting the full cosmological information from these datasets requires robust and scalable analysis methods that correct for systematic biases and model systematic uncertainties.  An important  systematic uncertainty, sourced by coherent galaxy alignments with their neighbors and the large scale structure, is the \textit{intrinsic alignment} (IA) of galaxies \citep[e.g., for a review see][]{troxel-ishak}. 
  This effect, if not accounted for, can falsely imitate weak gravitational lensing signals and bias cosmological analyses \citep{weak-lens}. 
Cosmological simulations are often employed to derive synthetic galaxy catalogs for testing, improving, and validating analysis pipelines and their robustness to systematic uncertainties \citep{buzzard-1, buzzard-2, euclid-flagship}; and in future analyses, they may be used to constrain cosmology via Simulation-Based Inference \citep[SBI;][]{des-lfi, hsc-sbi}.  
As future surveys will cover large areas of the sky and produce high-quality data, cosmological simulations of increasingly larger volumes and higher resolutions, along with realistic systematics, like intrinsic alignments,  are needed \citep{Vogelsberger-review}.

 Cosmological simulations with galaxy evolution  are computationally intractable at the large scales and high resolutions demanded by future surveys \citep{Vogelsberger-review}. 
 One common approach for producing synthetic galaxy catalogs is to run N-body simulations and then ``paint'' in the galaxies using semi-analytic models  \citep{sam-benson, sam-shark, sam-galaxies7020056}.   
 Though these methods are less computationally costly than running a full hydrodynamical simulation, they are still costly and introduce tunable parameters. The   multivariate distribution
describing various galaxy and dark matter (sub)halo properties is called the galaxy-halo connection. 
Even though halo occupation distribution (HOD) and  subhalo abundance matching (SHAM) models have been shown to  predict  galaxy mass, 
abundance and clustering with some success \citep{somerville-dave, hod-sham}, 
recent studies show that  the full galaxy-halo connection is highly non-linear and high dimensional  \citep[see, e.g.,][for a comprehensive review]{wechsler-tinker}.

Recent advances in machine learning have sparked a revolution in astrophysics and cosmology, offering promising avenues for tackling complex challenges in modeling the universe \citep[for a recent review, see][]{dawes-review}. 
Relevant previous attempts to harness machine learning for the problem of intrinsic alignments considered in this work  include simple and limited implementations with generative adversarial networks (GANs)  \citep{graphgan}; novel approaches with geometric deep learning methods are emerging to overcome the hurdles identified in that work \citep{Jagvaral_Lanusse_Mandelbaum_2024}. 
Geometric deep learning methods have garnered a lot of attention with their ability to handle complex datasets that do not fit a grid-like structure (e.g., images) or datasets that do not live in a Euclidean space  \citep{bronstein}. Geometric deep learning is an area of research that seeks to exploit symmetries and structures that are inherent in the data using mathematical graphs, group theory and other geometric concepts.  As such, these methods allow us to model galaxy intrinsic alignments and orientations, while respecting the inherent symmetry of the problem.

 At small scales, non-linear gravitational evolution and baryonic physics drive  the physics of galaxies,   
 making it difficult to model  and capture the dependencies and correlations among galaxy and dark matter subhalo properties  \citep{cdm_small_scale,small_scale}. Fortunately, graph neural networks (GNNs), based on mathematical graphs, are a natural way to model galaxies, given their ability to model sparsely distributed objects and capture correlations among neighbors \citep{gnn-guide-atomic}. Additionally,   galaxy orientations belong to the Lie group SO(3), requiring  these data  to be modeled on the corresponding manifold.   

 Recently, diffusion-based generative models have become the state-of-the-art for various generative tasks, such as images, audio, video and molecules \citep{diffusion-review}. 
Modeling distributions on the manifold of 3D rotations is, however, a non-trivial task, and to address
this problem we developed a new type of score-based diffusion model specifically for the SO(3) manifold in \cite{Jagvaral_Lanusse_Mandelbaum_2024}, by extending the Euclidean framework introduced in \cite{song2021}. Diffusion models are flexible in their ability to   model  data that live in various different spaces (e.g., scalars and manifold-valued data) compared to normalizing flows \citep{compare_generative}.  
Additionally, diffusion models are more stable compared to Generative Adversarial Networks 
\citep{compare_generative}. 

Based on these developments, we propose to build a conditional generative model for joint galaxy properties, using diffusion-based models in Euclidean space \citep{song_improved_2020} for scalars; and diffusion-based models in SO(3) \citep{Jagvaral_Lanusse_Mandelbaum_2024} for galaxy orientations.
Moreover, the diffusion model is implemented with  a GNN architecture that is used to
jointly model the correlated galaxy properties within halos with E(3) equivariance  \citep{egnn}   to exploit  global  symmetries. 
 In this regard, generative machine learning approaches combined with state-of-the-art geometric deep learning methods could serve as robust emulators for synthetic galaxy catalog production.   In particular, they could be used to populate realistic galaxies in large volume dark matter-only simulations.


The paper is organized as follows: 
in \S~\ref{sec:shear}  we briefly introduce weak lensing and IA.   In \S~\ref{Problem} we  describe the problem specifics. In \S~\ref{data_section}, we describe the cosmological simulation and the astrophysical observables. We describe our methodology in  \S~\ref{methods}. In \S~\ref{results} we present the results, and finally in \S~\ref{conc_section} we conclude our paper and propose future work.

\section{Gravitational weak lensing and Intrinsic Alignments}\label{sec:shear}



   Coherent galaxy shape distortions caused by gravitational lensing due to large-scale structure are referred to as cosmic shear \citep{Kilbinger}. In gravitational weak lensing studies, the observed galaxy shape distortions $\gamma = \gamma^I + \gamma^G$ are assumed to come from coherent intrinsic alignments due to localized physics such as tidal alignments (first term) and the true weak   lensing signal caused by structure along the line of sight (second term). 
   These shape distortions are directly and linearly related to the observed galaxy ellipticities, with the exact relation depending on the ellipticity definition \citep{2014ApJS..212....5M}.  
Assuming that the tidal fields at the formation of the galaxy set the intrinsic shapes of galaxies through a linear response \citep{catelan,hirata-seljak}, the linear alignment model  can be  expressed as  
            	\begin{align}\label{eq:gamma_ia}
            		\gamma^I=(\gamma_+^I,\gamma_\times^I) \propto
                    			 \left(\partial_x^2-\partial_y^2,\partial_x\partial_y\right)\Phi  
            	\end{align}
                 where  $ \Phi$ is the gravitational potential at the time of galaxy formation and  $(x, y)$ is an arbitrary coordinate system  on the plane of the sky \citep{ia-review2}. 
                 However, this assumption is expected to break down on small scales (below 2--6 $h^{-1}$Mpc) where non-linear gravitational effects  take  over, along with other small-scale effects such as baryonic physics \citep[for a recent study, see][]{des-y3-ia} 
                 and necessitates specialized models with high capacity and expressivity, such as the halo model and GNNs \citep{VanAlfen, graphgan}. 
                
                 The $\gamma^G$ term takes a similar form that differs in its geometric configuration and involves the matter overdensity along the line of sight; given the scope of our paper we focus only on the $\gamma^I$ term. For the interested reader, we refer to the reviews by \cite{troxel-ishak,2015SSRv..193....1J,ia-review2,2015SSRv..193..139K}.
                 In the literature, the IA formalism is mostly expressed in 2D, given the observational nature of cosmological measurements. For a 3D simulation box, we make similar assumption relating   tidal fields and intrinsic galaxy orientations, and condition the large scale alignments on tidal fields.

\section{Problem Statement}\label{Problem}


   In general, the problem of emulating galaxies in cosmological simulations can be formulated as a \textit{joint} conditional probability density estimation problem over \textit{correlated}   properties of different galaxies  $\bf{x}_\text{galaxy}$, that may be conditioned on some information about their local environment $\bf{Q}$:
   %
\begin{equation}\label{prob_statement}
\centering
     p(\bf{x}_\text{galaxy}| \bf{Q} ) 
\end{equation}
In practice, addressing this joint conditional density estimation problem with analytic methods has been shown to be challenging given the highly complicated physics that drive galaxy evolution \citep{wechsler-tinker}. Recently, there have been studies focused on tackling this problem with deep learning, particularly for scalar quantities \citep{harry-desmond, ben-moews}.
In addition, some non-scalar galaxy properties, such as orientation and position, obey symmetries, adding additional constraints and complexity to the problem. 
Graph neural networks are specifically designed to capture these joint correlated densities 
and can be made to respect the symmetries of the problem \citep{bronstein}.

The other aspect of the problem is learning how to model various regimes in cosmological distance scales (linear vs.\ non-linear scales). 
Here we try to capture the complex interconnections and dependencies in the non-linear scales of the universe 
using graph neural networks (GNNs), namely E(3) equivariant graph neural networks (E3GNN). In order to exploit the global Euclidean symmetry, we chose  the analytically imposed E3GNN from \cite{egnn}, as opposed to implementing the symmetry solely through data augmentation. 

Galaxies are not randomly distributed, but rather are organized into clusters, filaments, and voids, forming a rich network of connections. 
GNNs can leverage this network structure to learn representations of galaxies that incorporate information from their neighboring environments, though the exact specificity of the connections may be designed to fit the problem and the  computational resources. 
By iteratively aggregating information from neighboring nodes through message passing, GNNs can capture the complex spatial correlations and hierarchical relationships inherent in the cosmic web \citep{bronstein-2017, graphgan}.  
GNNs provide a powerful framework for representing and analyzing these intricate relationships by treating the cosmic large-scale structure as a set of \textit{graphs}, where  \textit{nodes} represent galaxies, and \textit{edges} encode interactions between them. 

As for the linear alignments on large  scales, we  leverage the gravitational tidal forces experienced by galaxies to condition their spatial orientations. By incorporating this information   into the input features of a neural network architecture,  it becomes possible to predict the statistical properties of  intrinsic alignments of galaxies with much less complexity and computational cost, as opposed to building one single graph for the whole galaxy population in the simulation volume \citep{graphgan}. 

 In this paper, we focus on the problem of \textit{intrinsic alignments} of galaxies, where we are learning and predicting galaxy orientations in 3D jointly with their sizes. To do so, we model DM subhalos and galaxies as triaxial ellipsoids.
In subsequent sections we will enumerate the specific quantities $\bf{x}_\text{galaxy}$ that we predict, and the set of $\bf{Q}$ values on which they are conditioned.

\section{    Methodology}\label{methods} 



\subsection{ Score-based Diffusion for SO(3) and Euclidean data   }
\textit{Diffusion generative models} have emerged as a powerful class of machine learning techniques for generating complex data distributions. These models operate by simulating a diffusion process that progressively adds noise to the data, transforming it into a simple, often Gaussian distribution. The generative process is then learned by training the model to reverse this diffusion, effectively denoising the data step-by-step to reconstruct the original distribution.  Diffusion generative models are particularly notable for their stability and ability to model diverse types of data, including images, audio, and 3D molecular structures \citep{diffusion-review}. These models offer a robust framework for generating high-fidelity synthetic data, making them very promising for applications in astrophysics, where they could potentially be used to simulate galaxy distributions, orientations, and other complex galactic physics.


We start with a generic stochastic differential equation (SDE) 
with a deterministic and a stochastic term:
\begin{equation}\label{eqn:forward_sde}
    \diff \bx = \mathbf{f}(\bx, t) \diff t + g(t)\diff \mathbf{w}
\end{equation}
where $\bx$ is a data vector, $t$ is time and  $\diff {\mathbf{w}}$ is a infinitesimal Brownian motion. 
As noted in \citet{ANDERSON-1982, de_bortoli_riemannian_2022}, under mild regularity conditions this SDE has a reverse diffusion process, defined by the reverse-time SDE:
\begin{equation}
    \mathrm{d} \mathbf{x} = [\mathbf{f}(\mathbf{x}, t) - g(t)^2  \nabla \log p_t(\mathbf{x})] \mathrm{d} t + g(t) \mathrm{d} \bar{\mathbf{w}},\label{eqn:backward_sde}
\end{equation}
where $\bar{\mathbf{w}}$ is a reversed-time Brownian motion and the \textit{score function}\footnote{  In the  literature there are two different definitions of this score function: when the derivative is taken with respect to the model parameters $\theta$: Fisher score --  $\nabla_{\theta} \log p(\mathbf{x})$; and when the derivative is taken with respect to the data $x$: Stein score --  $\nabla_{x} \log p(\mathbf{x})  $. In diffusion models, the Stein score function is almost always used, and we adopt it here as well.}   $\nabla \log p_t(\mathbf{x})  $  is the derivative of the logarithmic marginal density of the forward process at time $t$.  
If we denote the data distribution at $t=0$ as $\mathbf{x}(0) \sim p_\text{data}$  and   $p_t$ the marginal distribution of $\mathbf{x}(t)$ at time $t > 0$, 
$p_0=p_\text{data}$  then at final time $T$, $p_T$ will typically tend to a known target distribution.

Following \cite{song2019}, we choose $\mathbf{f}(\bm{x}, t) = 0$ and $g(t)=\sqrt{\frac{\diff \epsilon(t)}{\diff t}}$ for a given choice of noise schedule\footnote{The noise schedule is a sequence that controls how much noise is added to data during the diffusion process.} $\epsilon(t)$, 
which reduces Eq.~(\ref{eqn:forward_sde}) to:
\begin{equation}
    \diff \mathbf{x} =\sqrt{\frac{\diff \epsilon(t)}{\diff t}} \diff \mathbf{w} \;. \label{eqn:ve_sde}
\end{equation}

Next we adopt the variance exploding noise schedule and set the noise schedules for the SDE:
\begin{equation}\label{noise_sched}
    \sigma_i^2 = 2\epsilon(t_i) \\
    \sigma_\mathrm{min} = \sigma_\mathrm{1} < \sigma_\mathrm{i} \dots < \sigma_\mathrm{N} = \sigma_\mathrm{max}
\end{equation}
which is usually chosen to be a geometric sequence and $i$ denote different timesteps. 
Since we are dealing with data of multiple types (scalar vs.\ SO(3)), here our formalism slightly diverges from \cite{song2019}.
Given a dataset $\bx = \{\bx^S, \bx^M\}$, where $\bx^S$ denotes the scalar-valued data and  $\bx^M$ denotes SO(3) manifold-valued data, we can assign the  prior distributions for each:
\begin{align}
   & \mathcal{U}_\text{SO(3)} \, \text{Uniform noise for Manifold of SO(3)} \\
   & \mathcal{N}(0, 2\epsilon_\mathrm{max} \mathbf{I}) \, \text{Gaussian noise for Euclidean scalars}
\end{align}
 where $2\epsilon_\text{max} = 2\epsilon(t_N) = \sigma^2_\text{max}$.
Next, for scalar values we define the standard noise kernels with a Gaussian distribution, while the $\mathcal{IG}_\text{SO(3)}$ from \cite{nikolayev70, Matthies80} is used for the SO(3) data:





\begin{align}
& p_\epsilon(\tilde{\bx}^M | \bx^M) = \mathcal{IG}_\text{SO(3)}(\tilde{\bx}^M ; \bx^M, \epsilon) \hspace{0.3cm} \text{for} \, \bx^M, \tilde{\bx}^M \in \mathrm{SO(3)}
\\
& p_\epsilon(\tilde{\bx}^S | \bx^S) \hspace{0.22cm} = \mathcal{N} (\tilde{\bx}^S ; \bx^S, 2\epsilon \mathbf{I}) \hspace{0.8cm} \text{for} \, \bx^S, \tilde{\bx}^S \in \mathbb{R}^n 
\end{align}
where $n$ is the dimension of the scalars and $\tilde{\bx}(t)$ denote noised data at time $t=t_i$\footnote{For notational brevity, we omit the time dependence on $\tilde{\bx}(t)$ and just write $\tilde{\bx}$.}.  
In short, the Isotropic Gaussian Distribution on SO(3): $\mathcal{IG}_\text{SO(3)}$ is the heat kernel (solution to the diffusion equation on the $\text{SO(3)}$ manifold) that has properties that are well suited for our method, such as closure under convolution (for the noising process) and computational tractability (often heat kernels on non-Euclidean manifolds are infinite series). 
For the algebraic expressions and quick discussion about $\mathcal{IG}_\text{SO(3)}$ distribution please refer to \S 2.1 of \cite{Jagvaral_Lanusse_Mandelbaum_2024}. 

These definitions allow us to write down the noised data distribution for each data type:

\begin{equation}
    p^M_\epsilon(\tilde{\bx}^M) = \int_{\mathrm{SO(3)}} p^M_\text{data}(\bx^{  M}) p^M_\epsilon(\tilde{\bx}^M | \bx^M ) \diff \bx^M \;,
\end{equation} 
\begin{equation}
    p^S_\epsilon(\tilde{\bx}^S) = \int_{\mathbb{R}^n}  p^S_\text{data}(\bx^{  S}) p^S_\epsilon(\tilde{\bx}^S | \bx^S ) \diff \bx^S \;,
\end{equation} 
and correspond to $p_t$, the marginal distribution of the diffusion process at time $t$: $p_{\epsilon(t)} = p_t$.


This process is entirely defined as soon as the \textit{score function} of the marginal distribution at any intermediate time $t$, $\nabla \log p_{\epsilon(t)}$, is known.

Samples from $p_0$ can be obtained first by sampling:
\begin{equation}\label{eq:sampling}
  \bx=\begin{cases}
    \bx_T^M , & \text{at  $t=T \sim \mathcal{U}_{SO(3)}$}.\\
    \bx_T^S , & \text{at    $t=T \sim \mathcal{N}(0,2\epsilon_\mathrm{max} \mathbf{I})$}.
   
  \end{cases}
\end{equation}
 then  evolving them with Eq.~(\ref{eqn:backward_sde}) down to $t=0$. However, we still need the score function at intermediate times $t$, $\nabla \log p_{\epsilon(t)}$.

 \subsubsection{The (Stein) score function}
We need to approximate the score function of the form:
\begin{equation}\label{stein_score}
     \nabla_{x} \log p_\epsilon( \tilde{\mathbf{x}} | \mathbf{x}) 
\end{equation}
First, let us consider $\{ X_i \}_{i=0}^{3}$, an orthonormal basis of the tangent space $T_\mathbf{e}$SO(3). The directional derivative of the log density of the noise kernel $p_\epsilon(  \tilde{\bx}^M |\bx^M$) 
can be computed as:
\begin{equation}
    \nabla_{X_i} \log p_\epsilon( \tilde{\mathbf{x}}^M | \mathbf{x}^M) =  \left. \frac{\mathrm{d}}{\mathrm{d} s} \log p_\epsilon( \tilde{\mathbf{x}}^M \expm(s X_i) | \mathbf{x}^M) \right|_{s=0} \;,
\end{equation}
which can be computed in practice by automatic differentiation given the explicit approximation formulae for the $\mathcal{IG}_\text{SO(3)}$ distribution from \cite{Jagvaral_Lanusse_Mandelbaum_2024}. Thus,    the score computations for the SO(3) manifold valued data are tractable, compared to the general Riemannian manifolds.

Now for the scalar valued data, since we chose the noising kernel to be $\mathcal{N} (\tilde{\bx}^S ; \bx^S, 2\epsilon \mathbf{I})$,  
and noting   Eq.~(\ref{noise_sched}), 
we 
obtain the explicit form
\begin{equation}
\nabla_{x} \log p_\epsilon( \tilde{\mathbf{x}}^S | \mathbf{x}^S) = - \frac{\tilde{\bx}^S -\bx^S}{\sigma^2} = - \frac{\tilde{\bx}^S -\bx^S}{2\epsilon}. 
\end{equation}
Therefore, the score function from Eq.~(\ref{stein_score}) can be decomposed:

\begin{equation}
 \nabla_{x} \log p_\epsilon( \tilde{\mathbf{x}} | \mathbf{x}) = \begin{cases}
    \nabla_{X_i} \log p_\epsilon( \tilde{\mathbf{x}}^M | \mathbf{x} ) \, \\
    \nabla_{x} \log p_\epsilon( \tilde{\mathbf{x}}^S | \mathbf{x} ) 
  \end{cases}
\end{equation}



To match this derivative, we introduce a neural score estimator $s_\theta$: 
\begin{equation}
\nabla_{x} \log p_\epsilon( \tilde{\mathbf{x}} | \mathbf{x}) \approx s_\theta(\mathbf{x}, \epsilon) : (\text{SO(3)} \times \mathbb{R}^n  ) \times \mathbb{R}^{+ \star} \rightarrow \mathbb{R}^3 \times \mathbb{R}^n
\end{equation}
where $\mathbb{R^{+ \star}}$ denotes strictly positive numbers.  This can be trained directly under a conventional denoising score matching loss:
\begin{multline*}
    \mathcal{L}_{DSM} = \mathbb{E}_{p_\text{data}(\mathbf{x})} \mathbb{E}_{\epsilon \sim \mathcal{N}(0, \sigma_\epsilon^2)} \mathbb{E}_{p_{|\epsilon|}(\tilde{\mathbf{x}} | \mathbf{x} )} \\ \left[ |\epsilon| \ \parallel   s_\theta(\tilde{\mathbf{x}}, \epsilon) - \nabla_{x} \log p_{|\epsilon|}( \tilde{\mathbf{x}} | \mathbf{x}) \parallel_2^2 \right]
\end{multline*}
where we sample at training time random noise scales $\epsilon \sim \mathcal{N}(0, \sigma_\epsilon^2/2)$ similarly to \cite{song_improved_2020}, and we write $|\epsilon|$ to ensure the variance is positive. The minimum of this loss will be achieved for $s_\theta(\mathbf{x}, \epsilon) = \nabla \log p_{\epsilon}$. 

We then run it backwards starting from:
\begin{equation}
  \bx_{t=0}=\begin{cases}
    \bx^M_T    \text{  $ \sim \mathcal{U}_{SO(3)}$}.\\
    \bx^S_T    \text{  $ \sim \mathcal{N}(0,2\epsilon_\mathrm{max} \mathbf{I})$}.
    
  \end{cases}
\end{equation}
then using the learned score and the corresponding differential equation
\begin{equation} \label{diff-eq}
   \, \diff \mathbf{x}_t = -\frac{1}{2} \frac{\diff \epsilon_t}{\diff t} s_\theta(\mathbf{x}_t, \epsilon(t)) \diff t
\end{equation}
or in the decomposed form:
\begin{equation}
  \diff \mathbf{x}_t =- \frac{1}{2}  \frac{\diff \epsilon_t}{\diff t} \left\{\!\begin{aligned}
       s^M_\theta(\mathbf{x}_t, \epsilon_t)   \\
    s^S_\theta(\mathbf{x}_t, \epsilon_t)  
  \end{aligned}\right\}
  \diff t
\end{equation}
 
The network specifications for $s_\theta(\mathbf{x}_t, \epsilon(t))$ can be found in Appendix~\ref{network-specs}.

We note that integrating these equations require special care for quantities on the SO(3) manifold. Indeed, in addition to minimizing the integration error, the iterative solver must also ensure that the variable remains on the manifold at each integration step. We use in practice Heun's method to integrate this system, taking advantage of a special geometric flavor of the method for the SO(3) part of the signal. We direct the interested reader to Appendix~\ref{appendix:heuns} for the algorithm and to \citep{lie-group-book} for more information on geometric ODE solvers.

 
 

\subsection{E(3) GNN: 3D Euclidean group equivariant graph neural networks}


\textit{Graphs}  $\mathcal{G} = \big(\mathcal{V}, \mathcal{E}) $ are a mathematical structure composed of set of objects: nodes $v_i \in \mathcal{V}$, and their pair-wise connections -- edges $e_{ij} \in \mathcal{E}$. 
In graph-based data, each node is associated with features that can be broadly categorized into  \textit{scalar}  node features $\mathbf{h}_i \in \mathbb{R}^{n}$, 
which represent various properties such as mass or color; and  \textit{geometric}  node features $\mathbf{v}_i \in \mathbb{R}^{3}$, which may represent spatial information like position, orientation, velocity. We will closely follow the formalism and notation of \cite{egnn} and \cite{gnn-guide-atomic} -- thus  $\bx^S=\mathbf{h}$ and $\bx^M=\mathbf{v}$ to connect with notation from the previous subsection.     

\textit{Graph Neural Networks} (GNNs) are a class of neural networks specifically designed to operate on graph structures, making them particularly adept at capturing the relationships and interactions between nodes.  A GNN can be made \textit{geometric}, dubbed Geometric Graph Neural Networks, where each node $i$ possesses scalar and \textit{geometric} features/representations $\mathbf{h}^{l}_i$ and $\mathbf{v}^{l}_i$, respectively, at layer $l$.  Nodes exchange information through message passing, where messages are computed using a learnable function \textit{Msg}. 
These messages are then aggregated from the neighbors $\mathcal{N}_i$ of each node $i$ using a permutation-invariant operator $\oplus$, ensuring that the order of nodes does not affect the result. Finally, the features of each node are updated using its current representation and the aggregated messages via a learnable function \textit{Upd}. The general form of the message passing between nodes $i$ and $j$ at layer $l$: \, $\mathbf{m}^l_{ij}$, and the subsequent feature update mechanism on node $i$, are mathematically described by the following equations:
\begin{equation}\label{eqn:gnn_msg}
   \mathbf{m}^l_{ij} = \text{Msg}\Big(\mathbf{h}^{l}_i, \mathbf{h}^{l}_j, \mathbf{v}^{l}_i, \mathbf{v}^{l}_j, d_{ij} \Big) 
\end{equation}
%
\begin{equation}\label{eqn:gnn_upd}
\mathbf{h}^{l+1}_i  \, , \,  \mathbf{v}^{l+1}_i = \text{Upd} \Big(  \mathbf{h}^{l}_i, \mathbf{v}^{l}_i,  \bigoplus_{\substack{j \in \mathcal{N}_i }} ( \, \mathbf{m}^l_{ij})
\Big)
\end{equation}
with    $d_{ij} = || \mathbf{v}^{l}_i - \mathbf{v}^{l}_j ||_2$ denoting the Euclidean distance between nodes $i$ and $j$.
Further, in order to enforce E(3) equivariance which include symmetries of rotations, translations and reflections, we follow \cite{egnn, hoogeboom22a} and   chose the \textit{Msg} and \textit{Upd} function as follows:
\begin{equation}\label{eqn:ft_msg}
\mathbf{m}^{l}_{ij} =  \phi_e(\mathbf{h}^{l}_i, \mathbf{h}^{l}_i, d^2_{ij} )  
\end{equation}\begin{equation}\label{eqn:ft_update}
\mathbf{h}^{l+1}_i = \phi_h \Big(\mathbf{h}^{l}_i ; 
                                  \sum_{j \neq i} \widetilde{e}_{ij} \, \mathbf{m}^{l}_{ij}  \Big)
\end{equation}
\begin{equation}\label{eqn:vec_update}
\mathbf{v}^{l+1}_i =  \mathbf{v}^{l}_i + \sum_{j \neq i} \frac{\mathbf{v}^{l}_i - \mathbf{v}^{l}_j} {d_{ij}+1} \phi_v (\mathbf{h}^{l}_i, \mathbf{h}^{l}_j, d^2_{ij})
\end{equation}
 with  $ \widetilde{e}_{ij} = \phi_\text{inf} (\mathbf{m}^{l}_{ij} ) \in (0,1)$. 
 The various $\phi$ functions are implementations of the $Msg$ and $Upd$ mentioned above, which are small Multilayer Perceptrons 
 whose specifications can be found in the Appendix B of \cite{hoogeboom22a}. 
 


Eqs.~(\ref{eqn:ft_msg})-(\ref{eqn:vec_update}) need to be equivariant under the group transformations. Trivially, $\mathbf{m}_{ij}$ and $\mathbf{h}^{l+1}_i$ are invariant, since their inputs and outputs are scalars. For Eq.~(\ref{eqn:vec_update}), the second term has $(\mathbf{v}^{l}_i - \mathbf{v}^{l}_j)$ in which the two translation transformation will cancel each other. Next, since (roto-)reflections are   linear operators, we can factor out the (roto-)reflections outside the parenthesis of the second term. Finally, when we add the first term $\mathbf{v}^{l}_i$, we get a translation term and we can again factor out  the (roto-)reflections. Thus, the whole equation is equivariant under E(3) transformations; for details see Appendix~A of \cite{hoogeboom22a}. 

We can stack $L$ number of GNN layers to form the E3GNN architecture block, which   essentially transforms (non-linearly) the initial features, $\mathbf{h}^{l=0}$ and $\mathbf{v}^{l=0}$,  to obtain abstracted representations $\mathbf{h}^{l=1..L}$ and $\mathbf{v}^{l=1..L}$.  Typically, for molecular modeling, $L$ is on the order of 2--8; we chose $L=5$ for our study. Empirically, we found that $L<4$ were not expressive enough and $L>6$ were too slow to use in practice.

\subsection{Graph construction}\label{graph_const} 
 
To construct the graph for the cosmic web, we follow \cite{graphgan}. First, we  grouped all of the subhalos and galaxies based on their parent halo using their group membership ID from the halo finder. There exist a few choices for modeling proximity graph relations between the members in a group, such as the Gabriel graph (which is a subset of the Delaunay triangulation) and the radius nearest neighbor graph  (r-NNG).  These different graph types  differ by their
connectivity, i.e., they have different adjacency matrices \citep{proximity}. In this study we employ the r-NNG to model the connectivity of our graphs with a radius of 1 $h^{-1}$Mpc.

 Given a  galaxy catalog,  an undirected graph based on
the 3D positions is built by placing each galaxy on a {graph node}.   Then, each node will have a list of features such as halo mass, subhalo mass, central vs.\ satellite identification (binary column) and tidal fields smoothed on several scales.   
Then for a given group (i.e., within a halo) the graphs are connected using r-NNG with a radius of 1 Mpc/h. To build the graph connection,  the  nearest neighbors within a specified radius for a given node are connected via the  {undirected edges} with {signals} on the graphs representing the alignments and the scalar features.  

\section{Astrophysical data products and observables}\label{data_section}


\subsection{The Cosmological Simulation}
The IllustrisTNG suite of cosmological hydrodynamical simulations was run using the moving mesh code \textsc{Arepo} \citep{arepo}  in periodic boxes of $\sim$ 50, 100, 300 Mpc,  each with three different varying resolutions \citep{{ tng-bimodal,pillepich2018illustristng, Springel2017illustristng, Naiman2018illustristng, Marinacci2017illustristng,tng-publicdata}}. The simulations were initialized using cosmological parameters from Planck CMB measurements, assuming   a flat $\Lambda$CDM model \citep{planck2016}. Here, we chose the IllustrisTNG100 dataset due to its balance between large volume and high resolution.  Alongside magneto-hydrodynamic prescriptions of galaxy evolution,  the simulations also include  radiative gas dynamics; star formation and evolution; various forms of feedback from supernova and AGN  \citep[for more details, see][]{tng-methods,tng-agn}. The friends-of-friends \citep[FoF;][]{fof} and \textsc{subfind}  \citep{subfind} algorithms were employed to identify halos and subhalos in the simulations, respectively. 

We used the latest snapshot at $z=0$  for our analysis.  A mass cut of $ \log_{10}(M_*/M_\odot) \ge 9$ was enforced in order to obtain reliable galaxy shape measurements and to avoid resolution limitations \citep{gal_decomp}. The galaxy morphologies were determined based on a dynamical decomposition of the star particles into angular momentum-dominated and disperson-dominated components; for more details, see  \cite{IA-bulge-disc}.

\subsection{Tidal field}\label{appendix:TF}


In order to predict galaxy and DM halo shapes and orientations, we use the tidal field as a conditioning feature.  The tidal field is defined as  the Hessian of the gravitational potential $\Phi$:
\begin{equation}
    T_{ij}(\mathbf{r}) = \frac{\partial^2 \Phi(\mathbf{r}) } {\partial r_i\partial r_j},
    \label{eq:tidal_field}
\end{equation}
Using Poisson's equation, the  Fourier-space smoothed tidal field, expressed in terms of the overdensity field $\delta$, with a smoothing scale $\eta$ is:
\begin{equation}
     \hat{T}_{ij }(\mathbf{k}) = 4\pi G \bar{\rho} \frac{k_ik_j}{k^2} \hat\delta(\mathbf{k}) \;\; e^{-k^2 \eta^2 /2}.
     \label{eq:tidal_fourier_gaussian}
\end{equation}
The overdensity field was obtain by downsampling 1 million DM particles to reduce computational cost. 
The tidal field was evaluated at the position of each galaxy, using a cloud-in-cell window kernel to interpolate between the centers of the grid points,
with  $\eta$:= 1 $h^{-1}$Mpc   on a mesh of size $1024^3$, with cell sizes given as $L_\text{box}/1024=0.073~ h^{-1}$Mpc.

 \subsection{Shapes and Orientation of Halos and Galaxies}\label{shapes_methods}


 We use the  \textit{simple} (as opposed to the reduced or the iterative) mass quadrupole moments ${ I}_{ij}$ 
to measure the shapes of galaxies (DM (sub)halos) using star particles (DM particles):   
\begin{equation}
\label{eq:simple_IT}
{ I}_{ij} = \frac{\sum_n m_n {r_{ni} r_{nj}} }{\sum_n m_n}.
\end{equation}
the summation index $n$ runs over all  particles of a given type  
in a given galaxy, where  $m_n$ is the mass of the $n^{\rm th}$ particle and $r$
is the distance between the galaxy/subhalo centre of mass and the $n^{\rm th}$ particle, with $i$ and $j$ indexing the three spatial directions,   the $x,y,z$  Cartesian coordinates of the simulation box. 

 The three   eigenvectors   of ${ I}_{ij}$, defined   as $\mathbf{s}_\mu=\{s_{x,\mu},s_{y,\mu},s_{z,\mu}\}^\tau$ and $\mu \in \{a,b,c\}$ with eigenvalues $\omega_a, \omega_b ,\omega_c$, are related to
  the half-lengths of the principal axes of the ellipsoid  by: $a\propto\sqrt{\omega_a}$, $b\propto\sqrt{\omega_b}$, and $c\propto\sqrt{\omega_c}$, such that $a \geq b \geq c$. 
 
To compute the projected alignment signals, we need to use the 3D mass quadrupole moments to define 2D projected shapes. Following \cite{joachimi-2013, graphgan}, we can obtain the projected 2D ellipse as such:
\begin{equation}
(e_1,e_2) = \frac{(B_{xx}-B_{yy}, 2B_{xy})}{B_{xx} + B_{yy} + 2 \sqrt{\mathrm{det}\mathbf{B}}}.
\end{equation}

where 
\begin{equation}
\label{eq:ellipseprojection1}
{\mathbf B}^{-1} = \sum_{\mu=1}^3 \frac{ \bm{s}_{\perp,\mu} \vb \bm{s}_{\perp,\mu}^\tau}{\omega_\mu^2} - \frac{\bm{k} \bm{k}^\tau}{\alpha^2}\;,
\end{equation}
and
\begin{equation}
\label{eq:ellipseprojection2}
\mathbf{ k} = \sum_{\mu=1}^3 \frac{s_{\parallel,\mu} \bm{s}_{\perp,\mu}}{\omega_\mu^2}\;~~~\mbox{and}~~
\alpha^2 = \sum_{\mu=1}^3 \left( \frac{s_{\parallel,\mu}}{\omega_\mu} \right)^2\;.
\end{equation}
Here, $\mathbf{ s}_{\perp,\mu} = \{s_{x,\mu},s_{y,\mu}\}^\tau$ are the eigenvectors projected along the projection axis (for which we arbitrarily choose the $z$-axis of the 3D simulation box).  
In terms of the projected simulation box, the $x,y$ directions correspond to the positive and negative direction of $e_1$.


\subsection{Two-point estimators }
 Here we describe the two-point correlation functions that are used to quantify IA. The ellipticity-direction (ED) correlation captures the position and the orientation correlation angles in 3D,
 whereas the projected density-shape correlation function ($w_{g+}$) captures the correlation between overdensity and projected intrinsic ellipticity (which includes information on both shape and orientation in 2D).   All of the two-point statistic were measured using the  \href{https://halotools.readthedocs.io/en/latest/}{HALOTOOLS} package v0.7 \citep{halotools} and the supporting \href{https://github.com/duncandc/halotools_ia}{halotools\_ia} package .

\subsubsection{Density-Orientation Correlation Functions in 3D}

The ellipticity-direction (ED) correlation function is defined for each orientation axis $ \mu \in \{a,b,c\}$ as follows \citep{ed-ee}: 
\begin{equation}
\omega_\mu(r) = \langle |\mathbf{s}_\mu ({\bf x_\mathrm{pos}}) \cdot \vec{\mathbf{r}}({\bf x_\mathrm{pos}})|^2 \rangle -\frac{1}{3}=\omega_\mu^{1h}(r) + \omega_\mu^{2h}(r)
\end{equation}
for a subhalo/galaxy at position $\bf x_\mathrm{pos}$ with   axes direction $\mathbf{s}_\mu$ and the unit vector $\vec{\mathbf{r}}$  
denoting the direction of a density tracer at a distance $r$. 
Additionally,  this correlation function may be decomposed into the sum of  1-halo and 2-halo terms, where the 1-halo term captures the correlation among  subhalos within the same halo, and the 2-halo term captures contributions from pairs of subhalos belonging to different halos. 
 The estimator we use to compute it in the simulations is as follows:
 \begin{equation} 
   \omega_\mu (r) =  \sum_{\alpha \neq \beta} | \mathbf{s}_\mu^{\alpha} \cdot \vec{\mathbf{r}}^{\alpha \rightarrow \beta}|^2  - \frac{1}{3}
   \end{equation}
where  the orientation vector $\mathbf{s}_\mu^{\alpha}$ of  galaxy $\alpha$ is dotted with the unit vector pointing from the position of galaxy $\alpha$ galaxy to the position of galaxy $\beta$, $\vec{\mathbf{r}}^{\alpha \rightarrow \beta}$.

\subsubsection{Density-Shape Correlation Functions in 2D}
The cross correlation function of galaxy positions (or density tracers) and intrinsic ellipticities is defined as:
\begin{equation}
    \xi_{g+} (\mathbf{r} ) = \langle \delta_g(\mathbf{x}) \delta_+(\mathbf{x}+\mathbf{r}) \rangle
\end{equation}
where $ \delta_g(\mathbf{r})$ and $\delta_+(\mathbf{r})$ represent the galaxy overdensity field and the intrinsic shape field, respectively. It  can be estimated using the method described in \citet{mandelbaum-2011} 
as a function of $r_\mathrm{p}$ 
 (plane-of-sky distance) and $\Pi$ (line-of-sight distance): 
\begin{equation}
\xi_{g+} (r_\mathrm{p}, \Pi) = \frac{S_+D - S_+R}{RR}.
\end{equation}
 Here,  $RR$ are counts of  random-random  pairs  binned based on their perpendicular and line-of-sight separation; 
\begin{equation}
S_+D \equiv \frac{1}{2} \sum_{\alpha\neq \beta} e_{+}(\beta|\alpha),
\end{equation}
 represent the shape correlations,  
 where   $e_{+}(\beta|\alpha)$ is the $+$ component of the ellipticity of galaxy $\beta$ (from the shape sample) measured relative to the direction of galaxy $\alpha$ (from the density tracer sample). $S_+R$ is defined in an equivalent way, but instead of using galaxy positions we use randoms.

 The  two-point correlation functions can be projected onto 2D by integrating over the third dimension, with the integral approximated as sums over   the line-of-sight separation ($\Pi$) bins:
 \begin{equation}\label{eq:xi_to_w_sum}
w_{g+} (r_\mathrm{p}) = \sum_{-\Pi_\mathrm{max}}^{\Pi_\mathrm{max}} \Delta\Pi \,\xi_{g+} (r_\mathrm{p}, \Pi),
\end{equation}
where we chose a  $\Pi_\mathrm{max}$ value of 20 $h^{-1}$Mpc, following \cite{IA-bulge-disc}.

 \section{Results}\label{results}


 In this section, we present and investigate the outputs from the neural network. 
 The results presented here are from a single  random seed, since stochastic variation due to different random seeds has been shown to be subdominant to other sources of uncertainty \citep{graphgan,keller}. The train-test
splits were roughly 50/50, whilst preserving the halo membership
of galaxies. For orientation/alignment statistics and subsamples, we present the results on the whole dataset in order to increase signal-to-noise. 
 
First, for the scalar outputs, we will present the corner plot to illustrate the pairwise correlations.  Second, in order to quantify the correlations of geometric quantities we measure and present the 3D Ellipticity-Direction (ED) correlation functions.  Third, we present 2D projected quantities that are of great interest for weak gravitational lensing. We remind the reader of Eq.~(\ref{prob_statement}), and we list the inputs and outputs of the model in Table \ref{tab:in-out}.

The scalar features of interest for weak lensing science are the $a,b,c$ values that characterize galaxy/subhalo size, defined in \S\ref{shapes_methods}; or equivalently defined as $q=b/a, s=c/a$ on the interval $(0,1]$. In addition, we predicted   relevant scalars such as  galaxy mass, DM subhalo mass and galaxy color. In addition, the output geometric features are galaxy orientations and DM-subhalo orientations.
     

In principle, one can 
add as many scalar quantities as one wishes, however due to the \textit{neural scaling laws} the network capacity needs to be increased, in turn leading to a need for increased computer resources \citep{neural-scaling-laws}. 

\subsection{Investigations into generated galaxy features}

  \begin{figure*}
     \centering
         \includegraphics[width=7in]{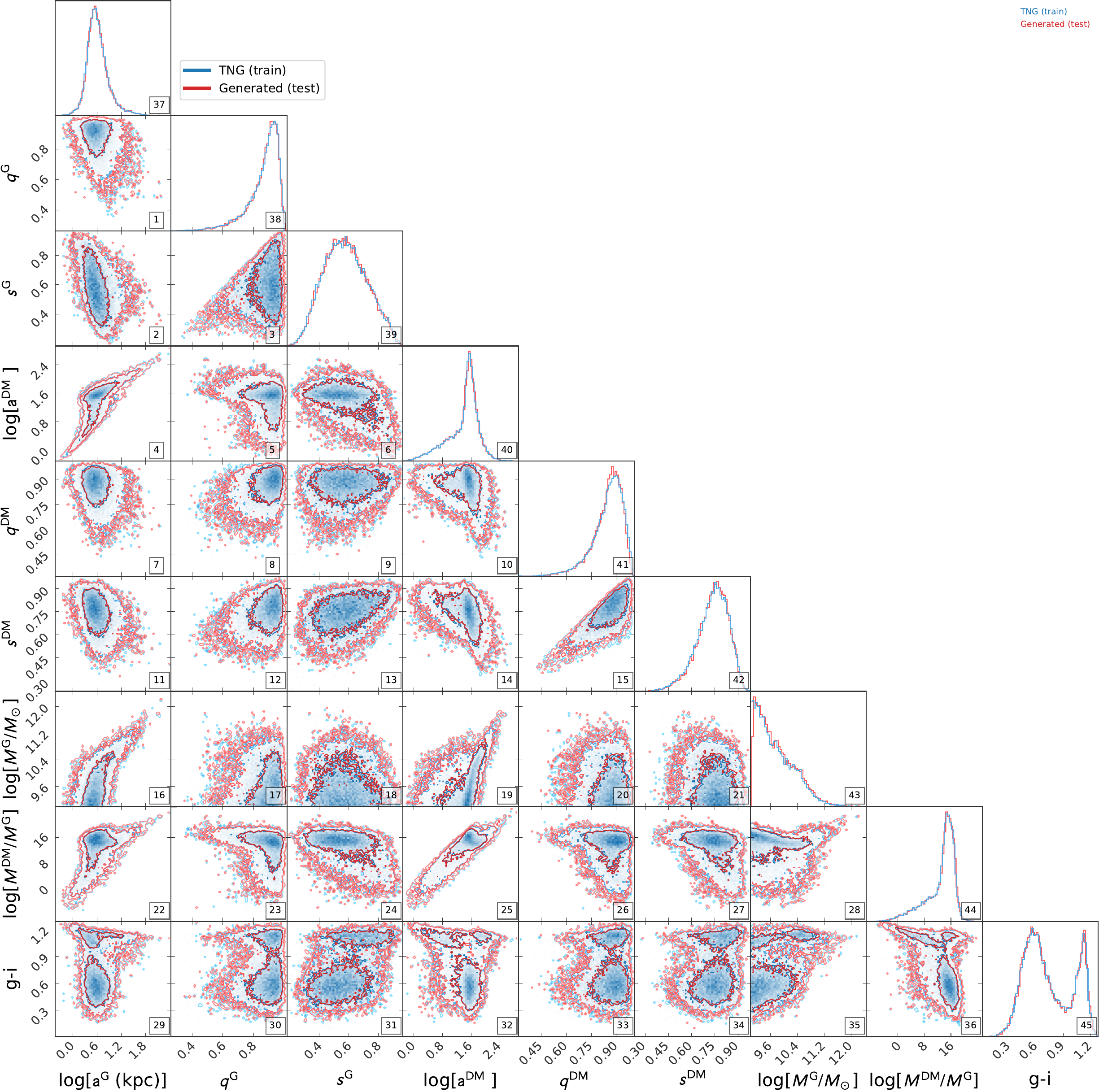}
 \caption{ \label{F:corner} Corner plot of correlations and 1D histograms between scalar quantities in the true TNG training sample versus the generated testing sample. The two joint distributions visually match well, and  quantitative metrics are  provided in Table~\ref{ks-table}. 
 }
\end{figure*}

\renewcommand{\arraystretch}{1.5}

\begin{table*}
    \centering
   \begin{tabular}{m{2.7cm} | m{1.2cm}  | m{2.8cm}  | m{7.8cm}} 

    \hline  \hline
     $\mathbf{Q} $ (Conditional Input)  &    Data-Type   & Constraints & Description   \\ [0.9ex] 
 \hline\hline

       { \large  $\vec{\bm{\lambda}_1}$, $\vec{\bm{\lambda}_2}$, $\vec{\bm{\lambda}_3}$}    &  $\mathbb{R}^3$ & ||$\vec{\bm{\lambda}_i}$||=1 &  eigenvectors of tidal field, defined in \S \ref{appendix:TF} (at subhalo positions)   \\   
        \hline
        
   { \large  ${{\lambda}_1}$, ${{\lambda}_2}$, ${{\lambda}_3}$}    &  Scalar &  --   &  eigenvalues of tidal field, defined in \S \ref{appendix:TF} (at subhalo positions)  \\   
 
 \hline
    { \large  $M_\text{total}$ }    & Scalar &  --   &   subhalo total mass: stellar+DM+gas  \\ 
    \hline
    { \large  $\leftindex^s{f}^c$ }    & Binary &  \{0,1\}   &   binary column indicating centrals vs. satellites  \\ 
    


 \hline  \hline
     $\bx $ (Modeled Quantities)  &        &   &     \\ [0.9ex] 
 \hline\hline

       { \large  log[$a^\mathrm{G}$ ] }    &  Scalar & (0, $\infty$) &  galaxy size:  semi-major axis length, defined in \S \ref{shapes_methods}   \\     
 \hline
   { \large  $q^\mathrm{G}$}   & Scalar &  (0, 1]  &   galaxy shape: intermediate-to-major axis ratio  \\ 
 \hline
    { \large  $s^\mathrm{G}$ }    & Scalar &  (0, 1]   &   galaxy shape: minor-to-major axis ratio  \\ 
 \hline
 { \large  log[$a^\mathrm{DM}$ ] }     & Scalar &  (0, $\infty$)  &   DM-subhalo size:  semi-major axis length, defined in \S \ref{shapes_methods}\\    
 \hline
   { \large  $q^\mathrm{DM}$ }  & Scalar &  (0, 1]  &  DM-subhalo shape: intermediate-to-major axis ratio     \\ 
 \hline
    { \large  $s^\mathrm{DM}$ }  &  Scalar &  (0, 1]  & DM-subhalo shape: minor-to-major axis ratio    \\ 
 \hline
    { \large log[$M^\mathrm{G}/M_\odot$] }  & Scalar &  (9, $\infty$) &    log(galaxy stellar mass)   \\ 
 \hline
     { \large  log[$M^\mathrm{DM}/M^\mathrm{G}$]}   & Scalar &  -- &    log of ratio of (DM-subhalo mass) to the  (galaxy stellar mass)   \\ 
 \hline   
 { \large  $g-i$}  &  Scalar &  --  &  galaxy color    \\ 
 \hline

  \hline
     { \large  $\bx_{t=0}^{M, \text{G}}$}  &  SO(3) &  $S^3$: unit 3-sphere  &  3D orientation of galaxy  \\ 
 
  \hline
     { \large  $\bx_{t=0}^{M, \text{DM}}$}  &  SO(3) &  $S^3$: unit 3-sphere  &  3D orientation of DM-subhalo \\
       \hline  \hline
    \end{tabular}
    \centering
    \caption{Summary of input/output features with their names, data-type, constraints and description.}
    \label{tab:in-out}
\end{table*}

\begin{table}

 \begin{tabular}{m{2.0cm} | m{0.6cm} m{0.9cm} | m{0.9cm} m{0.8cm} | m{0.8cm}  } 
 \hline
     $\bx^\text{S}$ (Scalars)&    KS   & $p$-value & $\frac{\text{Mean(Gen)}}{\text{Mean(TNG)}}$ &
     $\frac{\text{Var(Gen)}}{\text{Var(TNG)}}$ & $W_1$  \\ [0.9ex] 
 \hline\hline

 { \large  log[$a^\mathrm{G}$ ] }    &  $ 0.0077 $ &  $ 0.68 $ &  $ 1.00 $ &  $ 1.01 $ &  $ 0.0028 $  \\     
 \hline
   { \large  $q^\mathrm{G}$}   & $0.014 $ &  $ 0.07 $   &  $ 0.99 $ &  $ 1.04 $ &  $ 0.0032 $ \\ 
 \hline
    { \large  $s^\mathrm{G}$ }    & $ 0.0056 $ &  $ 0.94 $   &  $ 1.00 $  &  $ 1.01 $ &  $ 0.0014 $ \\ 
 \hline
 { \large  log[$a^\mathrm{DM}$ ] }     & $ 0.0062$ &  $ 0.88 $  &  $ 1.00 $ &  $ 1.00 $ &  $ 0.0034 $ \\    
 \hline
   { \large  $q^\mathrm{DM}$ }  & $ 0.028$ &  $ 0 $  &  $ 1.00 $ &  $ 1.04 $ &  $ 0.0035 $ \\ 
 \hline
    { \large  $s^\mathrm{DM}$ }  &  $ 0.018 $ &  $ 0.05 $  &  $ 0.99$  &  $ 1.02 $ &  $ 0.0033 $  \\ 
 \hline
    { \large log[$M^\mathrm{G}/M_\odot$] }  &  $ 0.052 $ &  $ 0$ &  $ 0.99 $ &  $ 0.99 $ &  $ 0.0049 $ \\ 
 \hline
     { \large  log[$M^\mathrm{DM}/M^\mathrm{G}$] } &  $ 0.0045 $ &  $ 0.99 $ &  $ 1.00$ &  $ 1.00 $ &  $ 0.0022 $  \\ 
 \hline   
 { \large  $g-i$}  &  $ 0.0075 $ &  $ 0.71 $ &  $ 1.00 $ &  $ 1.00 $ &  $ 0.0018 $ \\ 
 \hline

\end{tabular}
\caption{
Kolmogorov-Smirnov test statistics, the corresponding $p$-values; ratio of the means;  ratio of the variances; and the Wasserstein-1 distance between the true TNG   and generated scalar galaxy properties included within $\mathbf{x}_\text{galaxy}$, as defined in Eq.~(\ref{prob_statement}). The KS test $p$-values are consistent with the null hypothesis that the generated samples and the TNG samples are drawn from the same underlying distribution for unconstrained quantities, while constrained quantities such as $q, s$ and galaxy mass have $p$-values $\sim 0$.  The ratios of the first 2 moments are all $\sim 1$. The Wasserstein-1 distances of $\leq 10^{-2}$ also indicate the similarity of the distributions. 
}
\label{ks-table}
\end{table}

In Figure~\ref{F:corner}, we visualize and compare the joint distribution of scalar quantities using a corner plot of the generated and the TNG samples. Below the diagonal cells of the plot grid, the scatter plot along with contour lines helps us visualize the correlations across scalar quantities. On the primary diagonal, the marginal distributions of each scalar feature are shown.  As shown,  both the marginal and scatter/contours show good agreement between the generated and the TNG sample. 
Interestingly, the model has captured the bimodal distribution of galaxy colors as well as the unimodal distributions, such as the galaxy sizes/shapes.  
 Further, when broken down by mass, morphology and central/satellite type, the model has learned the trends with these quantities to a good degree, showing similar agreement between the two distributions (not shown here for brevity, but Figure \ref{F:wgp} implicitly shows this, since it uses three scalars to compute the projected 2D galaxy shapes).

In order to quantify the agreement, we calculate   Kolmogorov-Smirnov test statistics (KS-test); the ratio of the means; the ratio of the variances; and Wasserstein-1 distances  on these quantities and present them in Table~\ref{ks-table}.  
For all quantities, the KS-test results are all below $10^{-1}$; for $\{a^\text{G},s^\text{G},a^\text{DM}, M^\text{DM}/M^\text{G}, g-i\}$, the $p$-values  support   the null hypothesis that the two samples came from the same distribution. 
However, the corresponding $p$-values for the constrained quantities such as $q,s$ and mass are below the standard threshold of $0.05$. This may be due to the ``edge'' of the distribution shifting due to the   cutoffs imposed on these parameters, which are challenging for the neural network to pin down.  
However, in our intended astrophysical application, these low $p$-values may be negligible given that the shape  of the distributions match very well, as seen by the ratios $\sim 1$ of the mean and variance in Table~\ref{ks-table}. Lastly, we have computed the Wasserstein-1 ($W_1$) metric for these scalars and it shows distances below $10^{-2}$. 
 The $W_1$ distance measures the minimum cost required to move  one probability distribution to another, where the cost is the distance moved times the amount of ``probability distribution''. Compared to the KS-test, the $W_1$ metric is more sensitive to the overall shape of the distributions, whereas 
 for   large samples, even small   differences can lead to a rejection of the null hypothesis for a KS-test \citep{nonparametric-ks}. Based on these findings, we conclude that the generated scalars are robust.  Additionally, when broken down by mass, morphology and central/satellite type (not shown on Fig.~\ref{F:corner}), the scalar quantities exhibited similar agreement between TNG-100 and the generated samples, both visually and quantitatively.

  \begin{figure*}
     \centering
         \includegraphics[width=7in]{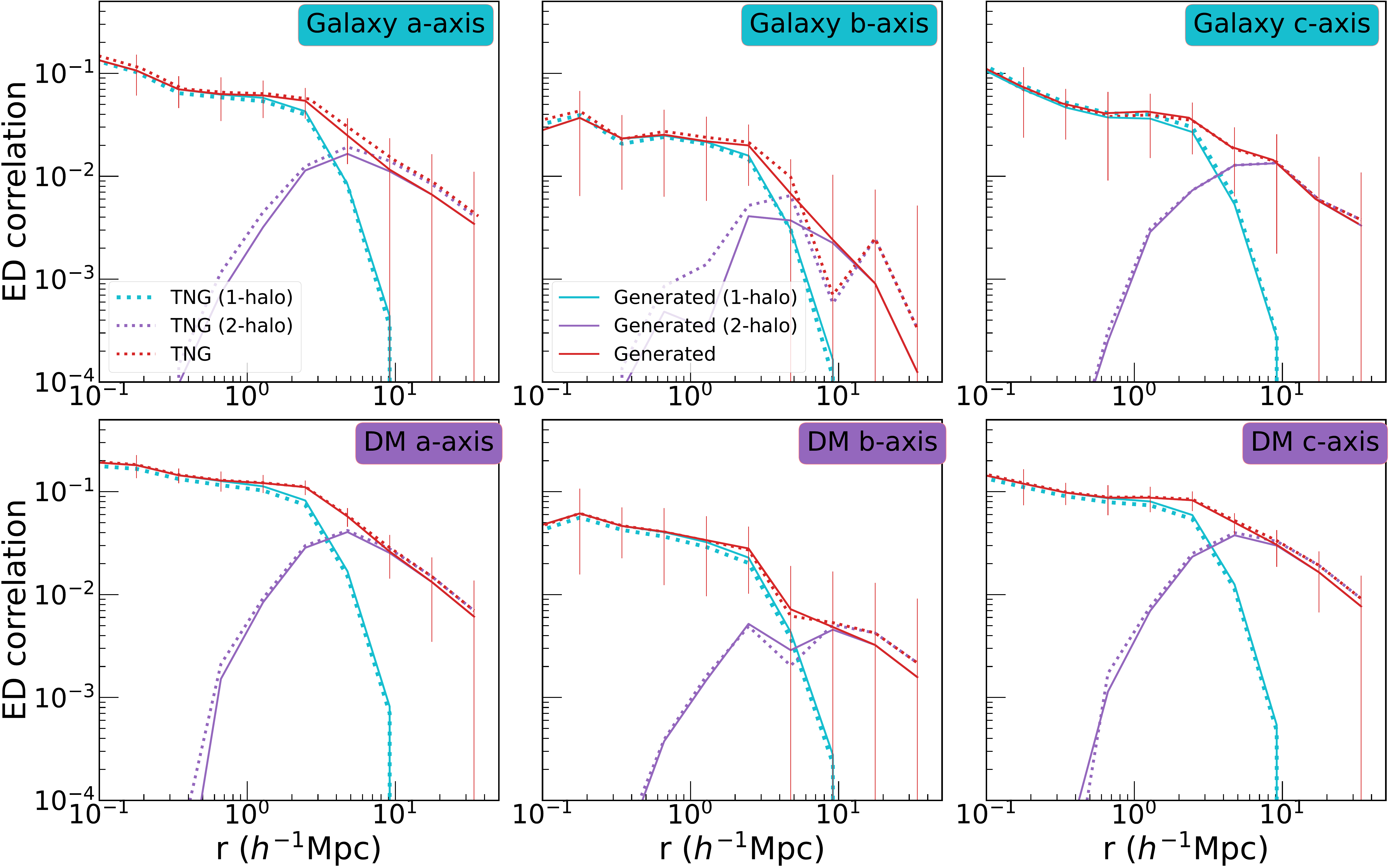}
 \caption{ \label{F:ED_all} Ellipticity-Direction (ED) correlation function  of galaxies (top row) and DM subhalos (bottom) for  the whole sample.  The ratio of the Generated ED function to the TNG ED function is shown in Figure~\ref{F:ED_ratio_reduced} along with the same ratio for  subsamples.
 The errorbars are shown for only 1 curve for visual clarity. We see an agreement between the true and generated values across all panels and on all scales. 
 }
\end{figure*}

   \begin{figure}
     \centering
         \includegraphics[width=3in]{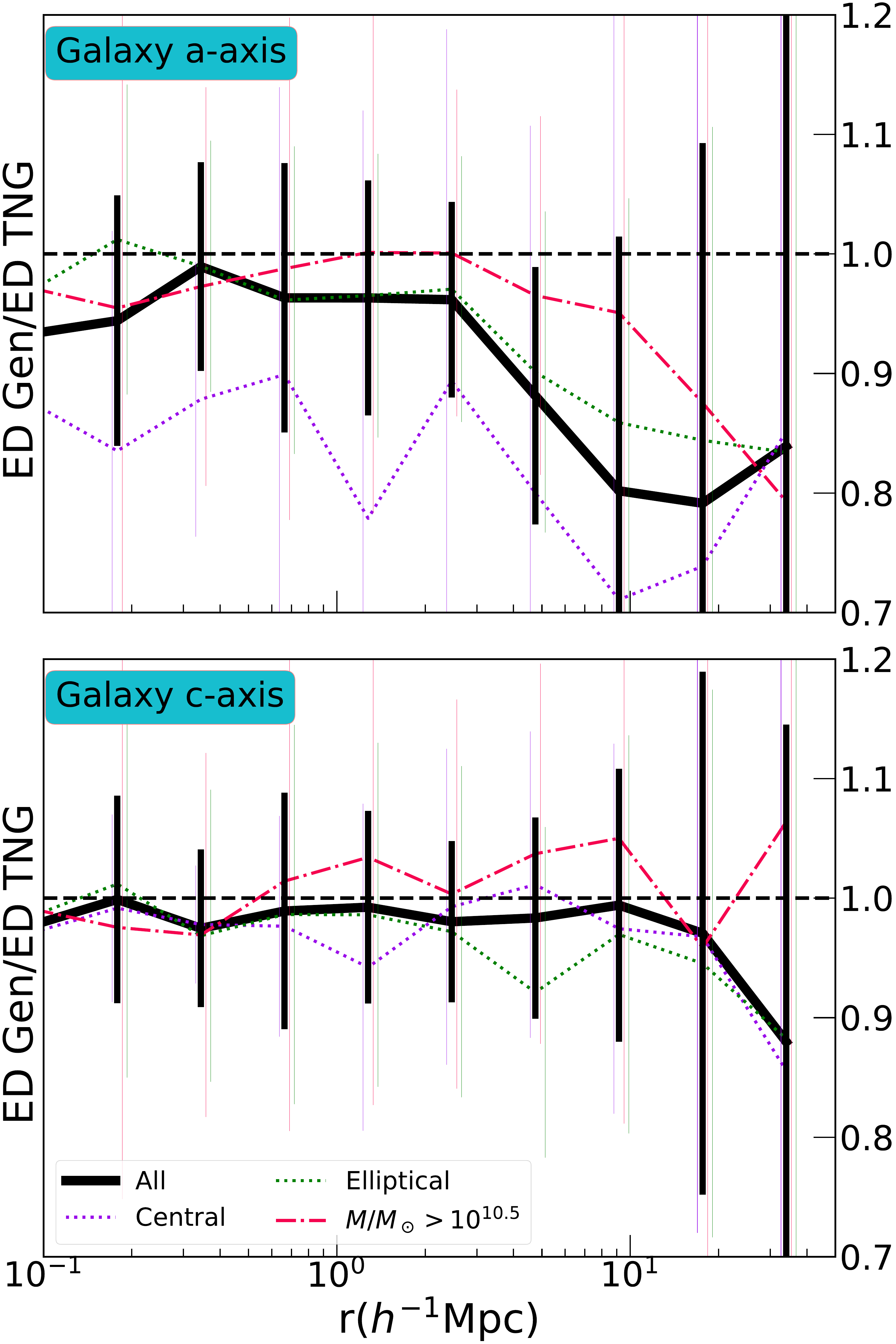}
 \caption{ \label{F:ED_ratio_reduced} Ratio of the Ellipticity-Direction (ED) correlation function  between the entire generated and TNG samples (black) and representative subsamples (colors indicated in legend).  The top (bottom) panel shows results for the longest (shortest) axis; we do not show results for the intermediate axis due to its very low signal.
 The errorbars on the subsample curves were horizontally shifted by 3\% for visual clarity. These ratio curves exhibit consistency with 1 given the statistical error of the measurements. 
 }
\end{figure}

Moving on to the geometric quantities, in Figure~\ref{F:ED_all} we present the ED correlations as a function of scale, as well as the 1- and 2-halo decomposed curves. 
The errorbars were obtained using jackknife estimates. 
The first (second) row contains the ED correlation function calculated using each of the 3 axes orientation of galaxies (DM subhalos) correlated with the density tracers. The generated samples reproduce the same correlated statistics as the TNG simulation across all scales within the given  statistical precision. We remind the reader that the 1-halo term is dominated by highly complicated physical interactions that  are hard to analytically model and the GNN has successfully handled this regime. While the 2-halo term is often assumed to be dominated by (linear) gravity, it still encompasses highly nonlinear scales,    and was   conditioned on the smoothed tidal field interpolated to the galaxy positions. This was done  to reduce the computational cost of the method. By \textit{not} connecting the halos, we save in the number of graph edges -- which dominates the computational expense.

When the shape sample is broken down by mass, morphology and the central-satellite classification, the results for the generated samples and TNG still agree well, as shown in Figure~\ref{F:ED_ratio_reduced} for representative subsamples.  Here we show the ratio of the ED curves of generated galaxy orientations against the true TNG galaxy orientations for the whole sample (black solid line) and for various subsamples (colored dashed/dotted lines). Given the very low signal of the intermediate axis and its non-prevalent use in literature, we omitted it for brevity, focusing only on the longest and shortest axis. 
The ratio  curves show  trends consistent with 1 given the errorbars. 
Of these curves, the centrals show trends furthest away from 1, probably due to our choice to build the graphs separately for each halo (thus centrals do not pass messages to each other during optimization via graphs). 
 The alignment of {centrals} is therefore mostly inferred from the tidal fields.  

Therefore, the learned model reproduces the target joint distributions to a good statistical agreement within our  error bars 
These results show us that geometric deep learning methods are a powerful way to model the highly complex joint probability density of the galaxy-halo connection. 
Compared with the GAN result from \cite{graphgan}, this model robustly and jointly captures all the features concurrently and in full 3D.

\subsection{Projected data products for weak gravitational lensing}

 \begin{figure*}
     \centering
         \includegraphics[width=7in]{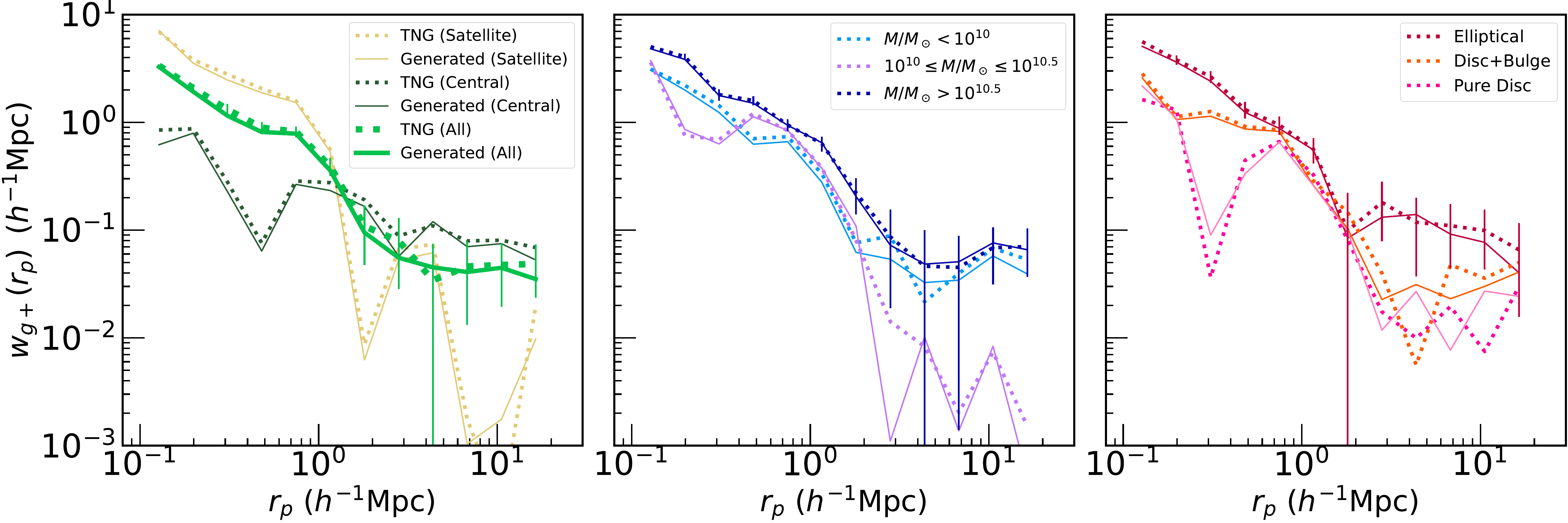}
 \caption{ \label{F:wgp} Projected two-point correlation functions $w_{g+}$ of \textit{all} galaxy positions and  projected 2D ellipticities of  \textit{(sub)samples}.  For visual clarity we only show the errorbars on a single measurement. 
 The left panel contains measurements for all galaxies   and for central and satellite  subsamples. The center (right) panel contains measurements for samples split by mass (morphology).
  The measurements with generated quantities show consistency across all scales with the true TNG measurements. 
 }
\end{figure*}

\begin{figure} 
            \includegraphics[width=84mm]{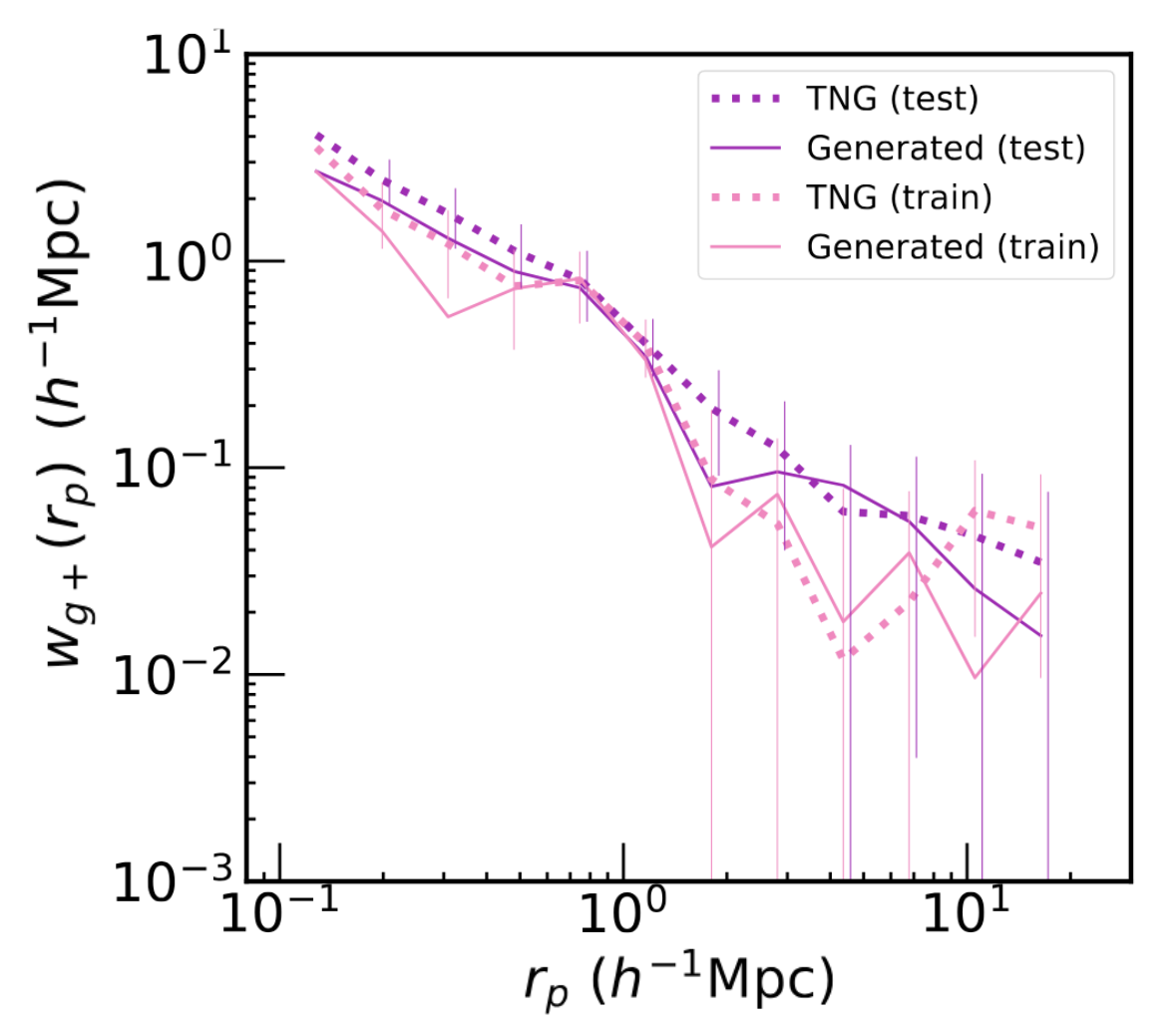}
        \caption{ Projected two-point correlation functions $w_{g+}$ of galaxy positions and the projected 2D ellipticities of train-test split samples.  The generated curves follow the TNG measurements closely, showing no signs of overfitting.
        }
        \label{F:train_test}
    \end{figure}

We project the 3D triaxial ellipsoidal model for the galaxy onto 2D in order to produce and compare the quantity that is typically measured to characterize the IA contamination to the weak gravitational lensing observables, namely the $w_{g+}$ correlation function.
We follow the   procedure described in \S~\ref{shapes_methods} to project the 3D inertia tensor onto 2D to obtain the complex ellipticities $e_1$ and $e_2$.
The measured $w_{g+}$ functions are presented in Figure~\ref{F:wgp}; in all of these curves the density tracers are the same -- all galaxies above our adopted mass threshold.  
The curves from generated samples follow the TNG curves within the statistical errors.  The sub-populations also exhibit good agreement with the truth. Notably, morphological information was not given to the model, yet the model learned the trends with morphology. 
Hence, the method we propose in this paper can potentially be used in current and future weak lensing surveys to generate realistically complex mock galaxy catalogs that can be used to test the analysis pipelines. 

In the application of machine learning to cosmology, one critical step involves avoiding overfitting, which is commonly done by dividing the data into training and testing sets. However, 
simulated data are limited by computational cost. In order to decrease the shot noise, we used the
full TNG sample as both the training set and testing set in Figures~\ref{F:ED_all}, \ref{F:ED_ratio_reduced}, and~\ref{F:wgp}. 
To test the impact of this choice and ascertain whether our results may be subject to overfitting, in Figure~\ref{F:train_test} we show the train  and test split samples for the $w_{g+}$ correlation function. The train-test splits were done roughly 50/50, preserving the group membership of galaxies by sampling the halos.
The  two TNG curves presented in   Figure~\ref{F:train_test} are more noisy than the curve presented in Figure~\ref{F:wgp} due to the decreased sample size, but they do agree within the errors. As for the $w_{g+}$ measurements from the generated samples, they follow very closely the true curve and are statistically consistent, if we assume the same covariance estimates. Therefore, we conclude that our model has not been subject to significant overfitting.

To summarize, our results demonstrate that the integration of machine learning techniques with geometric deep learning methodologies provides a robust framework for modeling complex joint probability distributions of galaxy properties that are governed by highly non-linear galactic physics. By effectively capturing both scalars and orientations, the model shows strong agreement with the true TNG samples across various scales, sub-populations and parameters. This advance underscores the potential of geometric deep learning to tackle complex problems in astrophysics, 
paving the way for more precise and tractable mock galaxy catalogs for testing and improving future cosmological analyses pipelines. One critical step that was not demonstrated in this paper is the performance of the model when various simulation suites are considered; we will explore this in future work.





\section{Conclusions} \label{conc_section}

We have presented a machine learning based method for modeling the galaxy-halo connection, specifically  focusing on the problem of intrinsic alignments of galaxies and their dependence on properties such as mass. Leveraging recent advances in geometric deep learning methods, we have modeled the galaxy distributions within dark matter halos as a set of graphs, where the nodes represent galaxies and edges representing their interaction, whilst incorporating the global E(3) symmetry. Additionally, the 3D intrinsic orientations  of galaxies were incorporated into a novel diffusion-based model for SO(3). 
We have demonstrated the efficacy of our method on the TNG100-1 hydrodynamical simulation from the IllustrisTNG suite with $\sim$18k galaxies. The generated samples from the trained model agree well with the true data, showing consistent correlations for both geometric and scalar quantities. 

This successful demonstration of our methodology motivates its extensions towards practical applications in future work.  
One issue to be addressed in future is that this model has custom-tuned, possibly overtuned, 
hyperparameters that may be specific to TNG. We do not know if these hyperparameters will give us similar quality results when training or testing on other simulations.  Extending this work to other simulations is a crucial step for future work.   Also, in future work, the model is planned to be deployed on a much lower resolution high-volume simulation than the one considered here.  Additional tests must be done on lower resolution simulations within the TNG suite to understand resolution effects on these models. 

Moving forward, it is essential to continue refining and expanding upon these machine learning methodologies to enhance the efficiency and robustness of mock galaxy catalog production, 
as well as to deploy these methods to test the data analysis pipelines for future cosmological surveys, such as Rubin LSST.  
Future studies could explore additional ML architectures and datasets, whilst incorporating more sophisticated machine learning techniques such as transfer learning \citep{TL, TL-astro} 
on various cosmological simulations. 
In an upcoming paper, we will investigate how to fully generate galaxy samples given a low resolution halo catalog, where we anticipate the outcome to be of comparable quality and realism as hydrodynamical simulations at a far lower computational cost. This in turn will allow an alternative and economical way of producing mock galaxy catalogs. 

\section*{Acknowledgements}

We thank Chad Schafer  and
Rupert Croft for useful feedback.  This work was supported in
part by 
a grant from the Simons Foundation (Simons Investigator in
Astrophysics, Award ID 620789).   This work
is supported by the NSF AI Institute: Physics of the Future, NSF
PHY-2020295. The computations in this work were, in part, run at facilities supported by the Scientific Computing
Core at the Flatiron Institute, a division of the Simons Foundation. YJ is thankful to Francisco Villaescusa-Navarro for hosting YJ at the Simons Foundation.


\section*{Data Availability}
The IllustrisTNG data can be obtained through the website at
\url{https://www.tng-project.org/data/}. The catalog data with morphological decompositions of galaxies is available at \url{https://github.com/McWilliamsCenter/gal_decomp_paper}.
The software developed as part of this work will be made available after peer-review.



\bibliographystyle{mnras}
\bibliography{example} 




\appendix

\section{Details of the neural network architecture}\label{network-specs}
The diffusion neural network has a size of \{4096, 2048, 1024, 512, 256, 128\} neurons each with GELU activation along with 3 layers of E(3) GNNs. 
We trained our models using the Adam optimizer with a learning rate  of $10^{-4}$, exponential decay rates of $\beta_1=0.90$ and $\beta_2=0.95$, 1 million   iterations, and then we decrease the learning rate to $10^{-5}$ for 0.5 million steps
with a batch size of 128. NVIDIA Tesla A100 GPU was used as the hardware, with JAX, Jraph and DeepMind-Haiku Python libraries as the software.

\section{Solving Differential equations: Heun's method}\label{appendix:heuns}


Heun's method, also known as the explicit trapezoidal rule, is a numerical method of solving initial value ODEs. 
\subsection{ Euclidean manifold case}
Given an ODE of the form:
\begin{equation}
    y'(t) = f(t,y(t)), \qquad \qquad y(t_0)=y_0, 
\end{equation}
with  timestep  $h$ and $t_{i+1}=t_{i}+h$, the approximate final value at the next integration point is given by:

\begin{align}
   & P_1 = h f(t_i, y_i) \\
   & P_2 = h f(t_i+\frac{1}{2}h, \,  \frac{1}{2}P_1 + y_i)  \\
   & y_{i+1} =  P_2  + y_i  
\end{align}

\subsection{  SO(3) manifold case }

Let us define $\phi$ to be the map from Lie algebra  $ \mathcal{G}$ to the Lie group $G$, for example the exponential map ($\phi  = \mathrm{expm}$), then we can write:

\begin{align}
   & F_1 = h f(t_i, y_i) \\
   & F_2 = h f\left(t_i+\frac{1}{2}h, \, \phi\left(\frac{1}{2}F_1\right) y_i\right)  \\
   & y_{i+1} = \phi(F_2) y_i  
\end{align}
These solutions can be obtained numerically for Eq.~(\ref{diff-eq}) for a given set of time steps $t_i = 0 \dots T$ \citep{lie-group-book}.  
We use equal timesteps with $T=1024$.




\bsp	
\label{lastpage}
\end{document}